\documentclass[showpacs,preprint,preprintnumbers,showpacs,showkeys,superscriptaddress,amsmath,amssymb,nofootinbib]{revtex4}

\usepackage[dvips]{graphicx}
\usepackage{color}
\usepackage{latexsym}
\usepackage{amsmath}
\usepackage{amssymb}
\usepackage{eufrak}
\usepackage{euscript}
\usepackage{pstricks}
\usepackage{graphics}
\usepackage{graphicx}
\usepackage{picture}
\def\[{\left\lbrack}
\def\]{\right\rbrack}

\def\({\left(}
\def\){\right)}

\newcommand{\bee}{\begin{equation}}
\newcommand{\eee}{\end{equation}}
\newcommand{\eaa}{\end{eqnarray}}
\newcommand{\baa}{\begin{eqnarray}}

\newcommand{\p}{\partial}
\def\ni{\noindent}

\begin{document}

\title{\Large Gauge-invariant extensions of the Proca model \\ in a noncommutative space-time}

\author{Everton M. C. Abreu}\email{evertonabreu@ufrrj.br}
\affiliation{Grupo de F\' isica Te\'orica e Matem\'atica F\' isica, Departamento de F\'{i}sica, Universidade Federal Rural do Rio de Janeiro, 23890-971, Serop\'edica - RJ, Brazil}
\affiliation{Departamento de F\'{i}sica, Universidade Federal de Juiz de Fora, 36036-330, Juiz de Fora - MG, Brazil}
\author{Jorge Ananias Neto}\email{jorge@fisica.ufjf.br}
\author{Rafael L. Fernandes}\email{rafael@fisica.ufjf.br}
\author{Albert C. R. Mendes}\email{albert@fisica.ufjf.br}
\affiliation{Departamento de F\'{i}sica, Universidade Federal de Juiz de Fora, 36036-330, Juiz de Fora - MG, Brazil}

\date{\today}

\begin{abstract}
\ni The gauge invariance analysis of theories described in noncommutative (NC) space-times can lead us to interesting results since noncommutativity is one of the possible paths to investigate quantum effects in classical theories such as general relativity, for example.   This theoretical possibility has motivated us to analyze the gauge invariance of the NC version of the Proca model, which is a second-class system, in Dirac's classification, since its classical formulation (commutative space-time) has its gauge invariance broken thanks to the mass term.  To obtain such gauge invariant model, we have used the gauge unfixing method to construct a first-class NC version of the Proca model.  We have also questioned if the gauge symmetries of NC theories, are affected necessarily or not by the NC parameter.  In this way, we have calculated its respective symmetries in a standard way via Poisson brackets.
\end{abstract}

\pacs{04.50.-h, 05.20.-y, 05.90.+m}
\keywords{Gauge unfixing formalism, gauge invariance, noncommutative Proca model}

\maketitle

\section{Introduction}

Nowadays it is common knowledge that gauge invariance is a fundamental stone in Standard Model theory. Consequently, the investigation of how to obtain models that are gauge invariant is an important procedure in several areas of research in theoretical physics.  

Another challenge in theoretical physics is how to connect quantum mechanics to general relativity.  It is believed to be the path to understand the physics of the early Universe for example \cite{NCreviews}.

Physical models described in noncommutative (NC) space-times are attempts to introduce a Planck measure, the NC parameter, into the model.  In this way we can analyze the consequences of the NC approach.  Namely, new terms arise due to the NC space-time and their influence in the physics of the systems can be discussed.

It is well known that gauge theories can be constructed in NC spaces by choosing systems (actions) that are invariant under gauge transformations defined through the well known Moyal-Weyl (MW) product \cite{sw} given by

\bee
\label{aaaA}
\hat{f}(\hat{x}) \star \hat{g}(\hat{x})\,=\, \hat{f}(\hat{x})\,\mbox{exp}\Big(\frac i2 \, \loarrow \partial^{x}_{\mu} \theta^{\mu\nu} \roarrow \partial^{y}_{\nu} \Big)\, \hat{g}(\hat{y}) \Big|_{x=y} \; ,
\eee
\noindent where  the hat notation means a NC space-time variable and $\theta^{\mu\nu}$ is the well known NC parameter that is present in the NC space-time definition, namely,
\bee
\label{bbbB}
\[ \hat{x}^{\mu}, \hat{x}^{\nu} \]\,=\,i\,\theta^{\mu\nu}\,\,,
\eee

\ni  where the space-time coordinates are promoted to the status of operators \cite{NCreviews}.  We can see clearly from (\ref{aaaA}) that if we consider higher $\theta$ terms in the MW product, it turns out to be a non-local product.  So, it is very common in the literature to consider only terms of first-order in $\theta$.  The ordinary products of the system will be substituted by the MW product and after that we have to introduce the so-called Seiberg-Witten (SW) map in order to obtain the NC terms (i.e., the NC contributions).  This will be clear here in the future.

Hence, the shape of the gauge transformations imply that the generators' algebra have to be closed under commutation and anti-commutation relations.  This is the reason why $U(N)$ is a common choice for the symmetry group for NC extensions of Yang-Mills theories instead of $SU(N)$. However, other symmetries structures can be considered too \cite{jmspw,bcpvz,bs,armoni,sheikh,af}.  Since a NC gauge theory has been constructed, one can find the SW map connecting the NC fields to the ordinary ones \cite{ak}.  The mapped Lagrangian is usually formulated as a nonlocal infinite series of ordinary fields.   However, their space-time derivatives and the NC Noether identities are kept by the SW map.  This guarantees that the mapped theory is still gauge invariant.

Having said that, we believe that something is missing concerning the analysis of gauge invariance of models described in NC space-times in terms of the constraints' (first or second-class in Dirac's formulation \cite{dirac}) analysis of the system.  The discussion of the constraints in order to classify them following Dirac's nomenclature, still has little attention in the NC literature.   In order to fill this gap we have discussed in this work the constraint analysis and the gauge invariance of the NC version of the Proca model, which was constructed in 
\cite{darabi}.  In this NC formulation of the Proca model, we will see that it is also a second-class system \cite{darabi} and in this way, to analyze gauge invariance, we have to convert the system into a first-class one, which is gauge invariant, as well known.

The method used here to convert second into a first-class system is the gauge unfixing (GU) method \cite{Vyt,jorge}, which was not used in any NC version of any well known model so far.  We will analyze its usefulness here under the presence of terms coupled to the NC parameter.   This is the second objective here since the first one is the result itself, namely, the obtention of a gauge invariant model concerning the Proca one.  The fact that the mass term breaks the gauge invariance in the standard commutative Proca model, turns the recovering of gauge invariance of its NC version an interesting result, since the terms connected to the NC parameter can be considered to have an effect analogous to the mass term in some other systems.

To accomplish the tasks that we are discussing in this introductory section, we have followed a sequence where in the next section, we have reviewed briefly the gauge unfixing method.   In section 3 we have described the NC version of the Proca model.  In section 4, the gauge invariant Proca model in NC space-time was obtained as well as the gauge symmetries.  The conclusions completes the paper and are written in the last part, section 5, together with some perspectives for futures research.


\section{A brief review of the gauge unfixing formalism}

Consider a second-class constrained system described by its correspondent Hamiltonian which has, for example, two second-class constraints $T_1$ and $T_2$. The basic idea of the GU formalism is to convert a second-class system into a first-class one by selecting one of the two second-class constraint to be the gauge symmetry generator, i.e., this constraint will be ``defined" {\it ad hoc} as being first-class. The other constraint will be discarded since a new first-class Hamiltonian will be constructed.  However, since we have two constraints, the next step is to build another conversion with the second constraint that was discarded.  Now this second constraint will be the chosen one, and the first constraint will be discarded.   To sum up, we have two cases in this GU formalism, namely, two ways to obtain gauge invariance.   This will be clear in section 4.

The idea is to interpret the original gauge-non-invariant theory as being a gauge-fixed version of the gauge invariant theory. If we choose $T_1$ as the generator of the symmetries then 
the second-class Hamiltonian must be modified in order to satisfy a first-class algebra. To accomplish that, the new and gauge invariant Hamiltonian can be constructed  through a series of powers of $T_2$ in order to not generate any more constraints, of course.  Hence, with this necessity in mind, we can write conveniently that

\begin{eqnarray}
\label{GH}
\tilde{H}=H+T_2 \{H,T_1\}+\frac{1}{2!} T_2^2 \{\{H,T_1\},T_1\}+\frac{1}{3!} T_2^3 \{\{\{H,T_1\},T_1\}T_1\}+ \ldots \,\,,
\end{eqnarray}

\ni where it can be shown that $\{\tilde{H},T_1\}=0$ (i.e., there are no secondary or any new constraints) and $T_1$ must satisfy a first-class algebra $\{T_1,T_1\}=0$.  In this way this final system was shown precisely to be a first-class, gauge invariant one.  

As we said before, one of our objectives here is to investigate the behavior of this formalism in the analysis of NC systems, since the NC contribution can bring difficulties, like a non-local first-class Hamiltonian, for example, since the MW product is extremely non-local at higher-orders of the NC parameter.  Although here we will limit the product to first-order terms of $\theta$, this possible non-locality could has its origin in the ``normalization" process described just above.  This fact could bring a theoretical impossibility which could make us to abandon this conversion method.  In other words, the GU conversion method could be considered incompatible with NC systems.


\section{Noncommutative Proca Model}

Let us review the main steps of the NC Proca model \cite{darabi}. The action of this model can be written as 
\begin{eqnarray}
\label{action}
{\cal S} = \int \( -\frac{1}{4} \hat{F}_{\mu\nu} * \hat{F}^{\mu\nu} +\frac{1}{2} m^2 \hat{A}_\mu*\hat{A}^\mu\) d^4x,
\end{eqnarray}\ni 

\ni where $*$ means the Moyal-Weyl product, $\hat{A}_\mu$ and $\hat{F}_{\mu\nu}$ are the vector potential and field strength tensor respectively described in the NC space-time (remember that it is the meaning of the hat notation), $m$ is the mass of the $A_\mu$ field and we will use $(-+++)$.  We can notice that in Eq. (\ref{action}) the NC terms do not appear since the fields involved live in the NC space-time.  To make the NC terms explicit, as we said before, the fields have to be rewritten in terms of the commutative fields through the SW map \cite{sw}, connecting both the commutative and NC terms.
Using the SW map, the fields $\hat{A}_\mu$ and $\hat{F}_{\mu\nu}$ can be written in terms of the corresponding commutative quantities as \cite{Bi}

\begin{eqnarray}
\label{fmn}
\hat{F}_{\mu\nu}=\partial_\mu\hat{A}_\nu -\partial_\mu\hat{A}_\mu -i \hat{A}_\mu*\hat{A}_\nu + i \hat{A}_\nu*\hat{A}_\mu,
\end{eqnarray}

\ni where
\begin{eqnarray}
\label{amn}
\hat{A}_\mu=A_\mu-\frac{1}{2} \theta^{\alpha\beta} A_\alpha (\partial_\beta A_\mu+F_{\beta\mu}),
\end{eqnarray}

\ni and the $\theta^{\alpha\beta}$ term is the NC contribution.  One basic property in NC theory \cite{NCreviews} is that the integral over the star product of two quantities is equal to the corresponding integral over the ordinary product \cite{Riad}, leading the action (\ref{action}) be rewritten in the form 

\begin{eqnarray}
\label{actionn}
{\cal S} = \int \( -\frac{1}{4} \hat{F}_{\mu\nu}\hat{F}^{\mu\nu} +\frac{1}{2} m^2 \hat{A}_\mu\hat{A}^\mu\) d^4x \,\,,
\end{eqnarray}

\ni and the next step is to use the SW map written in (\ref{amn}). Hence, by using Eqs. (\ref{aaaA}), (\ref{fmn}) and (\ref{amn}), the NC density 
Lagrangian in (\ref{actionn}) can be written in terms of commutative fields as

\begin{eqnarray}
\label{lagn}
\hat{\cal L}= -\frac{1}{4} F_{\mu\nu}^2 + \frac{1}{8} \theta^{\alpha\beta} F_{\alpha\beta} F^2_{\mu\nu}-\frac{1}{2} \theta^{\alpha\beta} F_{\mu\alpha} F_{\nu\beta}
F^{\mu\nu} 
+\frac{1}{2} m^2 \Big[A^2_\mu-\theta^{\alpha\beta} A_\alpha(\partial_\beta A_\mu + F_{\beta\mu}) A^\mu \Big] \,\,,
\end{eqnarray}

\ni where we can see easily that the standard Proca model is recovered when $\theta=0$,
the commutative field strength tensor is the standard  $F_{\mu\nu}=\partial_\mu A_\nu-\partial_\nu A_\mu$.   Using the notation in \cite{darabi}, we can define the quantities

\begin{eqnarray}
A_\mu&=&(\vec{A},i A_0)\,\,,\nonumber\\
E_i&=&i F_{i4}\,\,,\nonumber\\
B_i&=&\frac{1}{2} \epsilon_{ijk} F_{jk}\,\,,\nonumber\\
\theta_i&=&\frac{1}{2} \epsilon_{ijk} \theta_{jk}\,\,,
\end{eqnarray}

\ni where the last one defines a vector carrying the NC parameter $\theta$ and $\p / \p t = i \p_4$.  Using these last definitions, the NC density  Lagrangian (\ref{lagn}) can be written in the following explicit form

\begin{eqnarray}
\label{10}
\hat{\cal L}&=&\frac{1}{2} (E^2-B^2) (1+\vec{\theta}\cdot\vec{B})-(\vec{\theta}\cdot\vec{E})(\vec{E}\cdot\vec{B})+\frac{m^2}{2} (-A_0^2+A^2) \nonumber\\
&+& \frac{m^2}{4} (\vec{\theta}\times \vec{A})\cdot\nabla(A_0^2)-\frac{m^2}{2} \Big[(\vec{\theta}\times\vec{A})\cdot\vec{E} \Big] A_0 \nonumber\\ 
&+& \frac{3}{4} m^2 \Big[(\vec{\theta}\cdot\vec{B})A^2-(\vec{\theta}\cdot\vec{A})(\vec{A}\cdot\vec{B})\Big] \,\,,
\end{eqnarray}

\ni where $A^2\,=\,\vec{A} \cdot \vec{A}$, obviously.  We can see clearly, as it is expected in NC space-time theories, the introduction of the NC parameter broke the explicit Lorentz invariance.  But the discussion of this topic is out of our scope.
Since our objective here is to convert second into first-class constraints, it is convenient to study the dynamics of the NC system described above in the scenario of the Hamiltonian framework.  In this way, following \cite{darabi}, the next standard step is to calculate the  momenta conjugated to $A_0$ and $A_i$ respectively 

\begin{eqnarray}
\label{pi0}
\pi_0&=&\frac{\partial\cal{L}}{\partial(\partial_0 A_0)}= 0,\\
\label{pii}
\vec{\pi}&=&\frac{\partial\cal{L}}{\partial(\partial_0 A_i)}\,=\,-\,(1+\vec{\theta}\cdot\vec{B})\,\vec{E}+(\vec{\theta}\cdot\vec{E})\,\vec{B}+(\vec{E}\cdot\vec{B})\,\vec{\theta}+\frac{m^2}{2} (\vec{\theta}\times \vec{A})\, A_0 \,\,.
\end{eqnarray}

\ni Following Dirac's constraint formulation, we can see clearly that Eq. (\ref{pi0}) is the primary constraint
\begin{eqnarray}
\label{fi1}
\phi_1\equiv\pi_0\approx 0\,\,.
\end{eqnarray}

\ni Using the Legendre transformation and Eq. (\ref{pii}) to re-express $E_i$ in terms of $\pi_i$ and first order in $\theta$, we obtain

\begin{eqnarray}
{\cal{H}}_c &=&\frac{1}{2} (\pi^2+B^2)+ \frac{1}{2} (B^2-\pi^2) (\vec{\theta}\cdot\vec{B})
+ (\vec{\pi}\cdot\vec{\theta}) (\vec{B}\cdot \vec{\pi})+ \frac{m^2}{2} A_0^2 \nonumber\\
&-&\frac{m^2}{2}\,A_0 [(\vec{\theta}\times\vec{A})\cdot \vec{\pi}]  -\frac{m^2}{2} \vec{A}^2 (1+\frac{3}{2} \vec{\theta}\cdot \vec{B})
-\frac{m^2}{2} A_0 [(\vec{\theta}\times\vec{A}) \cdot (\nabla A_0)] \nonumber\\
&+&\frac{3 m^2}{4} (\vec{\theta}\cdot\vec{A}) (\vec{A}\cdot\vec{B})-(\vec{\pi}\cdot \nabla) A_0
+{\cal{O}}(\theta^2).
\end{eqnarray}

\ni Requiring the time persistence of the primary constraint (\ref{fi1}) 
\begin{eqnarray}
\label{brak}
\phi_2=\{\phi_1,{\cal H}_c\},
\end{eqnarray}

\ni where $H_c=\int {\cal{H}}_c d^3x$, we have the secondary constraint in first order $\theta$
\begin{eqnarray}
\label{fi2}
\phi_2\equiv\nabla \cdot \vec{\pi}+m^2A_0+\frac{m^2}{2} \nabla\cdot(\vec{\theta}\times\vec{A})A_0+\frac{m^2}{2}(\vec{\theta}\times\vec{A})\cdot\vec{E}
\approx 0 \,\,,
\end{eqnarray}

\ni which can be seen as the generalized form of the Gauss law in NC space.  The next step is to classify the constraints in first or second-class.  The result of the Poisson bracket between $\phi_1$ and $\phi_2$ is \cite{darabi}
$$
\Big\{\phi_1(x),\,\phi_2(y)\,\Big\}\,=\,-\,m^2\, \Big( 1\,+\,\frac 12 \nabla \cdot (\vec{\theta} \times \vec{A}) \Big)\, \delta (x-y) \,\,,
$$

\ni which means that both constraints are second-class and the NC system is not gauge invariant.  These two constraints define a surface in the NC phase-space. 

We can see this non-invariance in Eq. (\ref{lagn}) at once, since if we consider the standard local gauge transformation for the Maxwell electromagnetic theory, i.e., $\delta A_\mu(x) = \partial_\mu \epsilon(x)$, where $\epsilon(x)$ is the local gauge parameter, we have that $\delta F_{\mu\nu}=0$.  This invariance of $F_{\mu\nu}$ makes the Lagrangian in (\ref{lagn}) a gauge invariant one if $m=0$.   Hence, as in the standard case, the mass term breaks the gauge invariance of the model.  Notice that the NC term does not spoil the gauge invariance of the $m=0$ action, of course.
In the next section we will see that the GU method recovers the gauge invariance of (\ref{lagn}) by converting the second into first-class, the constraints calculated here.



\section{Gauge-invariant extensions of NC Proca model}

As we said before, our objectives here are two-fold, on one hand we will analyze the gauge invariance of a well known non-invariant theory like the Proca model, where the mass term broke the gauge invariance.  On the other hand we want to investigate, in first order of $\theta$-terms, if the GU formalism is able to recover the gauge invariance of such NC theory since, as we said before, the NC terms can bring a non-locality in the constraint renormalization.  
We believe that our results can add some light in the gauge invariance issue concerning theories described in NC phase-spaces.

It is important to notice that, by construction, as it can be seen in Eq. (2), noncommutativity spoils the explicit Lorentz symmetry since one direction turned out to be well established as defined in Eq. (2).  In this work this fact is not relevant since the objective here is the gauge invariance issue {\it per se}.  Other topics like unitarity or renormalization are not discussed here.

We will describe the two cases, which is a feature of the Proca model having two second class constraints, for the construction of gauge invariant Lagrangians, namely, we will construct two Hamiltonians and the respective Lagrangians, that are gauge invariant, following GU formalism.  Each case, will be characterized by the choice of one of the constraints as being the first-class one, i.e., the generator of the gauge symmetries.  The other constraint will be discarded.  After that, we will calculate the gauge transformations for each case.

\subsection{Case 1:}

We begin the GU formalism applied to NC Proca model in (\ref{lagn}) by redefining the constraints, Eqs.(\ref{fi1}) and (\ref{fi2}), as
\begin{eqnarray}
\label{chi}
\chi&=&\,\phi_1,\\
\label{psi}
\psi&=&\frac{\phi_2}{m^2 \Big[1+\frac{1}{2} \nabla\cdot(\vec{\theta}\times\vec{A})\Big]}\,=\,\frac{\phi_2}{m^2} \Big(1\,+\,\frac 12 \vec{\theta}\cdot \vec{B} \Big)\,\,,
\end{eqnarray}

\ni where we have used that $\theta \ll 1 $ and $\chi$ plays the role of $\tilde{T}$ defined in section 2.  Hence, the denominator of $\psi$ 
is the Poisson bracket of $\phi_1$ and $\phi_2$ and, $\psi$ is the ``normalized" constraint redefined through GU point of view.

The constraints $\chi$ and $\psi$ form a canonical conjugate pair, i.e.,
\begin{eqnarray}
\Big\{\chi(x),\psi(y)\Big\}=\,-\,\delta^{(3)}(x-y).
\end{eqnarray}

In the first case, following the GU method, we will consider that  $\chi$ in Eq. (\ref{chi}) will be the chosen first-class constraint and $\psi$ in Eq. (\ref{psi}) will be discarded. Then, using $\chi$ as
the gauge symmetry generator, we can compute the gauge-invariant Hamiltonian as being
\begin{eqnarray}
\label{fh1}
\tilde{\cal H}={\cal H}_c-\frac{1}{2m^2} \Big[1+\frac{1}{2} \nabla\cdot (\vec{\theta}\times\vec{A}) \Big]\psi^2\,\,.
\end{eqnarray}

\ni Notice that, substituting Eq. (\ref{psi}) into Eq. (\ref{fh1}) we can write the second term of Eq. (\ref{fh1}) as

\begin{eqnarray}
&&\frac{1}{2m^2} \Big[1+\frac{1}{2}\nabla\cdot (\vec{\theta}\times\vec{A}) \Big]\psi^2 \nonumber \\
&=&\frac{1}{2m^2} \Big(1+\frac{1}{2}\nabla\cdot(\vec{\theta} \times \vec{A}) \Big)\bigglb[(\nabla \cdot \vec{\pi})^2+2m^2(\nabla \cdot \vec{\pi}) A_0+(m^2A_0)^2+m^2 (\nabla \cdot \vec{\pi})[\nabla\cdot(\vec{\theta}\times\vec{A})]A_0 \nonumber\\
&-&m^2[(\vec{\theta}\times\vec{A})\cdot\vec{\pi}\,]\nabla \cdot \vec{\pi}+(m^2A_0)^2[\nabla\cdot(\vec{\theta}\times\vec{A})]-m^4\,A_0[(\vec{\theta}\times\vec{A})\cdot\vec{\pi}\,]\,\biggrb]\,\,.
\end{eqnarray}

\ni Then, substituting this last result into Eq. (\ref{fh1}),  we can rewrite the first-class Hamiltonian, Eq. (\ref{fh1}), as
\begin{align}
\label{fh}
\tilde{\cal H}_{1st\:case}\,&=\frac{1}{2}(\pi^2+B^2)+\frac{1}{2}(B^2-\pi^2)\vec{\theta}\cdot\vec{B}+(\vec{\pi}\cdot\vec{\theta})(\vec{\pi}\cdot\vec{B}) \nonumber\\
&-\frac{m^2}{2}\vec{A}^2(1+\frac{3}{2}\vec{\theta}\cdot\vec{B})+\frac{3}{4}m^2(\vec{\theta}\cdot\vec{A})(\vec{A}\cdot\vec{B}) \nonumber\\
&-\frac{1}{2m^2}(\nabla \cdot\vec{\pi})^2\,\Big[1\,-\,\frac 12\,\nabla\cdot(\vec{\theta}\times\vec{A})\Big]
\,+\,\frac 12 (\nabla\cdot\vec{\pi})\,(\vec{\theta}\times\vec{A})\cdot \vec{\pi}
\end{align}

\ni where we have used again the $\theta \ll 1$ approximation in order to eliminate the denominators with $\theta$-terms and the $A_0$ term was eliminated naturally. 

After some algebra, it can be verified that
\begin{eqnarray}
\Big\{\tilde{\cal H}_{1st\:case}(x),\chi(y)\Big\}=0 \,\,,
\end{eqnarray}

\ni which means that the Hamiltonian is gauge invariant.  
Hence, $\chi$ and $\tilde{\cal H}_{1st\:case}$ describe a consistent gauge theory in NC phase-space.   It can be shown that, although $\tilde{\cal H}_{1st\:case}$ is gauge invariant, it encompasses gauge noninvariant fields.


The dynamics of the NC gauge invariant (first-class) Proca model can be given in the usual way by
\begin{align}
\dot{A}_0=&\{A_0,\tilde{\cal H}_{1st\:case}\}_{PB} \,\,, \nonumber\\
\dot{\pi}_0=&\{\pi_0,\tilde{\cal H}_{1st\:case}\}_{PB} \,\,\,,\nonumber \\
\dot{\vec{A}}=&\{\vec{A},\tilde{\cal H}_{1st\:case}\}_{PB} \,\,,\nonumber\\
\dot{\vec{\pi}}=&\{\vec{\pi},\tilde{\cal H}_{1st\:case}\}_{PB} \,\,,
\end{align}

\ni which give us the values
\begin{align}
\dot{A_0}&=0 \,\,, \nonumber \\
\dot{\pi}_0&=0 \,\,,\nonumber\\
\dot{\vec{A}}&=\vec{\pi}-\nabla\,A_0-(\vec{\theta}\cdot\vec{B})\vec{\pi}+(\vec{\theta}\cdot\vec{\pi})\vec{B}+(\vec{B}\cdot\vec{\pi})\vec{\theta}
-\frac{m^2}{2}A_0(\vec{\theta}\times\vec{A}) \nonumber \\
&+\frac{1}{m^2}\nabla(\nabla\cdot\vec{\pi})
-\frac{1}{2}\nabla[(\vec{\theta}\times\vec{A})\cdot\vec{\pi}]\,\,,
\end{align}
\begin{align}
\dot{\vec{\pi}}&=-\nabla\times \Bigglb\{\vec{B}(1+\vec{\theta}\cdot\vec{B})+\frac{1}{2}\vec{\theta}(B^2-\pi^2)+\vec{\pi}(\vec{\pi}\cdot\vec{\theta})
-\frac{3}{4}m^2 \bigglb[\Big( \vec{A}^2+\frac{1}{3}A_0 \Big)\vec{\theta}\,  \nonumber \\
&+(\vec{A}\cdot\vec{\theta})\vec{A}\,\biggrb]\,\Biggrb\}+m^2\vec{A}\, \Big[1+\frac{3}{2}(\vec{B}\cdot\vec{\theta})\Big]-\frac{3}{4}\,m^2\,\Big[\vec{\theta}\,(\vec{A}\cdot\vec{B})+(\vec{\theta}\cdot\vec{A})\vec{B}\,\Big] \nonumber\\
&+\frac{1}{2}m^2A_0^2\,(\vec{\pi}\times\vec{\theta})\,-\,\frac 12(\nabla \cdot \vec{\pi})\,(\vec{\pi}\times \vec{\theta})\,-\,\frac{1}{4m^2}\,\vec{\theta}\times \nabla(\nabla\cdot \vec{\pi})
\end{align}

\ni where this dynamics is important in the new constrained surface defined only by $\chi$.

The gauge invariant Lagrangian can be written as 
\begin{align}
\label{lagfirstcase}
{\cal L}_{1st\:case}&=-\pi_0\dot{A}_0+\vec{\pi}\cdot\dot{\vec{A}}-\tilde{\cal H}_{1st\:case} \,\,,\nonumber\\
\end{align}

\ni where the minus sign of the $\dot{A}_0$ term is due to the metric. We can see, after some algebra, that Eq. (\ref{lagfirstcase}) can be written in a compact way as
\begin{align}
{\cal L}_{1st\:case}&=\hat{\cal L}+\frac{1}{2}m^2{A_0}^2-\frac{1}{2m^2}(\nabla\cdot\vec{E})^2\,\left(1+\frac 32(\vec{\theta}\cdot\vec{B})\right)\nonumber\\
&+\,(\nabla\cdot\vec{E})\left[\frac{1}{m^2}\vec{B}\cdot\nabla(\vec{\theta}\cdot\vec{E})+\frac{1}{m^2}\vec{\theta}\cdot\nabla(\vec{E}\cdot\vec{B})+\frac{1}{2}\nabla\cdot[(\vec{\theta}\times\vec{A})A_0]\,\right]\nonumber\\
&+\,A_0\left[(\nabla\cdot\vec{E})(1+\vec{\theta}\cdot\vec{B})+\vec{B}\cdot\nabla(\vec{\theta}\cdot\vec{E})+\vec{\theta}\cdot\nabla(\vec{B}\cdot\vec{E})+\frac{m^2}{2}\nabla\cdot[(\vec{\theta}\times\vec{A})A_0]\right] \nonumber \\
&-\frac{1}{2}\vec{E}\cdot\nabla[(\vec{\theta}\times\vec{A})\cdot\vec{E}]\nonumber\\
&-\,\frac{1}{2}(\nabla\cdot\vec{E})[(\vec{\theta}\times\vec{A})\cdot\vec{E}]+\frac{1}{4}m^2A_0^2\,[\nabla\cdot(\vec{\theta}\times\vec{A})]
+\frac{1}{2}[(\vec{\theta}\times\vec{A})\cdot\vec{E}](\nabla\cdot\vec{E}) \nonumber \\
&+\frac{m^2}{2}[(\vec{\theta}\times\vec{A})\cdot\vec{E}]\,A_0 \,\,,
\end{align}

\ni where $\hat{\cal L}$ is written in Eq. (\ref{10}).   It can be shown that the involved important physical quantities are gauge invariant under the gauge transformations generated by $\chi$.   We will be back to this subject in the future.  Of course, when $\theta =0$, we recover the commutative phase-space results.  It is worth to mention again that the presence of the $\theta$-vector implies that the Lorentz invariance is broken, as can be seen in the Lagrangian above.

\bigskip

\ni {\bf Gauge symmetries:}

\bigskip

It is well known from the analysis of classical constrained systems that the local symmetries can be constructed with the help of the first-class constraint \cite{alexei}.  It represents the Lagrangian counterpart of the canonical transformation generated by the constraint that lives on the phase space structure. 
Since we have seen that the Hamiltonian constraints could be found during the Dirac procedure, this provides a standard method for obtaining the symmetries.  Following this recipe, the gauge transformation can be given by
\bee
\label{zzzZ}
\delta \phi(\vec{x})\,=\,\epsilon(\vec{y}) \Big\{\phi(\vec{x}), T(\vec{y}) \Big\} \,\,,
\eee

\ni where $\phi$ is the variable which gauge transformation we are looking for, $\epsilon$ is the local gauge parameter  and $T$ is the first-class constraint. 

In both cases analyzed here we have a constant NC parameter and two second-class constraints.  The primary phase space is given by $(A_0, \pi_0, A_i, \pi_i)$, the chosen first-class constraint is given by $\chi$ and, following GU procedure, the other constraint, $\psi$, was discarded.  The gauge invariant Lagrangian, of course, was calculated for each case, i.e., for each first-class constraint and this constraint is the symmetry generator of the model.   Substituting these quantities in (\ref{zzzZ}), we have, respectively, the gauge transformations given by

\baa
\label{zZ}
\delta_{1st\:case} A_0(\vec{x}) &=& \epsilon(\vec{x}) \qquad\qquad \mbox{and}  \qquad\qquad 
\delta_{1st\:case} \vec{A}\,=\,0  \,\,,
\eaa

\ni which is the trivial gauge transformations for the Lagrangian in Eq. (\ref{lagfirstcase}).  It can be verified using vectorial identities and neglecting the  total derivative, the gauge invariance is proved. 

\subsection{Case 2:}

In the second case, we will choose
\begin{eqnarray}
{\chi}=\,\frac{1}{m^2}\,\Biglb\{\nabla \cdot \vec{\pi}+m^2\,\Big[A_0+\frac{1}{2}[\nabla\cdot(\vec{\theta}\times\vec{A})]\,A_0
-\frac{1}{2}\,(\vec{\theta}\times\vec{A})\cdot\vec{\pi}\Big]\Bigrb\} \,\,,
\end{eqnarray}

\ni as the first class constraint. The constraint 
\begin{eqnarray}
\psi\,=\,-\,\pi_0,
\end{eqnarray}

\ni will be discarded. A new constrained surface is defined different from the previous case.  The new gauge theory will be defined on this new surface.
Following the GU technique, the gauge invariant Hamiltonian is given by
\begin{eqnarray}
&&\tilde{\cal H}_{2nd\:case} \nonumber \\
&=&\frac{1}{2}(\pi^2+B^2)+\frac{1}{2}(B^2-\pi^2)\vec{\theta}\cdot\vec{B}+(\vec{\pi}\cdot\vec{\theta})(\vec{\pi}\cdot\vec{B})+\frac{m^2}{2}\,A_0\,\Big(\,A_0-(\vec{\theta}\times\vec{A})\cdot\vec{\pi}\Big) \nonumber\\
&-&\frac{m^2}{2}A^2(1+\frac{3}{2}\vec{\theta}\cdot\vec{B})-\frac{m^2}{2}[(\vec{\theta}\times\vec{A})\cdot(\nabla\,A_0]A_0+\frac 34\,m^2(\vec{\theta}\cdot\vec{A})(\vec{A}\cdot\vec{B})-\vec{\pi} \cdot\nabla\,A_0 \nonumber\\
&-&\psi\Biglb\{\nabla\cdot\Big[ \vec{A}\Big(1+\frac{3}{2}\vec{\theta}\cdot\vec{B}\Big)
+\frac{3}{4}[ \vec{\theta}(\vec{A}\cdot\vec{B})+ \vec{B}(\vec{\theta}\cdot\vec{A})]\Big]
+\frac{3}{2}[\nabla\cdot(\vec{\theta}\times\vec{\pi})]\,A_0 \nonumber\\
&+&\frac{1}{2}[(\vec{\theta}\times\vec{\pi})\cdot\nabla]\,A_0\Bigrb\}-\frac{1}{2m^2}\Big[(\nabla\psi)^2\Big(1+\frac{3}{2}\vec{\theta}\cdot\vec{B}\Big)+\frac{3}{2}(\vec{\theta}\cdot\nabla\psi)(\vec{B}\cdot\nabla\psi)\Big]\,\,,
\end{eqnarray}

\ni and we can verify that
\begin{eqnarray}
\Big\{\tilde{\cal H}_{2nd\:case}(x),\chi(y)\Big\}=0\,\,,
\end{eqnarray}

\ni which confirms the gauge invariance of $\tilde{\cal H}_{2nd\:case}$.

Consequently, the equations of motion are given by 
\begin{align}
\label{dynamics2}
\dot{\pi_0}&=\nabla\cdot\vec{\pi}-\nabla\cdot(\vec{\theta}\times\vec{\pi})-m^2A_0+\frac{m^2}{2}(\vec{\theta}\times\vec{A})\cdot\vec{\pi}-\frac{m^2}{2}A_0\,[\nabla\cdot(\vec{\theta}\times\vec{A})] \,\,,\nonumber\\
\dot{A_0}&=\nabla\cdot\Big[ \vec{A}\Big(1+\frac{3}{2}\vec{\theta}\cdot\vec{B}\Big)
-\frac{3}{2}[ \vec{\theta}(\vec{A}\cdot\vec{B})+ \vec{B}(\vec{\theta}\cdot\vec{A})]\Big]-\frac{3}{2}[\nabla\cdot(\vec{\theta}\times\vec{\pi})]\,A_0\nonumber\\
&-\frac{1}{2}[(\vec{\theta}\times\vec{\pi})\cdot\nabla]\,A_0+\frac{1}{m^2}\nabla \cdot \,\Big[(\nabla\psi)\Big(1+\frac{3}{2}(\vec{\theta}\cdot\vec{B})\Big)\Big]+\frac{3}{4m^2}\nabla\cdot\,\Big[\vec{\theta}(\vec{B}\cdot\nabla)\psi+\vec{B}(\vec{\theta}\cdot\nabla)\psi\Big] \,\,, \nonumber
\end{align}
\begin{align}
\dot{\vec{A}}&=\vec{\pi}-(\vec{\theta}\cdot\vec{B})\vec{\pi}+(\vec{B}\cdot\vec{\pi})\vec{\theta}+(\vec{\theta}\cdot\vec{\pi})\vec{B}
-\frac{m^2}{2}A_0\,(\vec{\theta}\times\vec{A})-\nabla A_0 \,\,, \nonumber\\
\dot{\vec{\pi}} &=-{\nabla}\times \bigglb\{\vec{B}(1+\vec{\theta}\cdot\vec{B})+\frac{1}{2}\vec{\theta}(B^2-\pi^2)+\vec{\pi}(\vec{\pi}\cdot\vec{\theta})-\frac{3}{4}m^2\,\Big[ \,(\,A^2+\frac{1}{3}\,A_0\,)\,\vec{\theta}+(\vec{A}\cdot\vec{\theta})\,\vec{A}\Big]\,\biggrb\}  \nonumber \\
&+m^2\vec{A}\Big(1+\frac{3}{2}(\vec{B}\cdot\vec{\theta})\Big)
+\frac{1}{2}m^2\,A_0^2(\vec{\pi}\times\vec{\theta})-\frac{3}{4}m^2\Big(\vec{\theta}(\vec{A}\cdot\vec{B})+(\vec{\theta}\cdot\vec{A})\vec{B}\Big) \nonumber \\
&-{\nabla}\psi-\frac{3}{2}{\nabla}\Big(\psi(\vec{\theta}\cdot\vec{B})\Big) 
+\frac{3}{2}\psi{\nabla}(\vec{\theta}\cdot\vec{B})+\frac{3}{4}[\nabla\cdot({\psi}\vec{B})]\vec{\theta}
+\frac{3}{4}(\vec{\theta}\cdot{\nabla})(\psi\vec{B})-\frac{3}{4}{\psi}(\vec{\theta}\cdot{\nabla})\,\vec{B}   \nonumber\\
&-\frac{3}{2}\vec{\theta}\times{\nabla}(\psi{\nabla}\cdot\vec{A})
-\frac{3}{2}\vec{\theta}\times{\nabla}({\nabla}\cdot\vec{A})+\frac{3}{4}{\nabla}(\vec{A}\cdot\vec{\theta})\times{\nabla}\psi
\nonumber \\
&-\frac{3}{4}\Big((\vec{\theta}\cdot{\nabla})(\vec{A}\times{\nabla})\psi+\psi{\nabla}\times[(\vec{\theta}\cdot{\nabla})\vec{A}\,]\Big) 
+\frac{3}{4m^2}\,\Big[\vec{\theta}\times{\nabla}({\nabla}\psi)^2+{\nabla}\psi\times{\nabla}(\vec{\theta}\cdot{\nabla}\psi)\, \Big]\,\,. \nonumber \\
\end{align}

The first-class Lagrangian can be obtained by carrying out the Legendre transformation and the resulting the first-class Lagrangian for the second case is
\begin{align}
\label{zzZ}
&\tilde{\cal L}_{2nd\:case} \nonumber \\
&=-\pi_0\dot{A}_0- \vec{\pi} \cdot \dot{\vec{A}}-\tilde{\cal H}_{2nd\:case} \nonumber \\
&=\tilde{\cal L}-\pi_0\partial_{0}A_0-\pi_0\bigglb\{\nabla\cdot \Big[\vec{A}\,\Big(1+\frac{3}{2}\vec{\theta}\cdot\vec{B}\Big)-\frac{3}{2}\Big(\vec{\theta}(\vec{A}\cdot\vec{B})+\vec{B}(\vec{\theta}\cdot\vec{A})\Big)\Big]
-\frac{3}{2}[\nabla\cdot(\vec{\theta}\times\vec{E})]\,A_0\nonumber\\
&-\frac{1}{2}\,[(\vec{\theta}\times\vec{E})\cdot\nabla]\,A_0\biggrb\}
+\frac{1}{2m^2}\Bigglb[(\nabla\pi_0)^2\Big(1+\frac{3}{2}\vec{\theta}\cdot\vec{B}\Big)+\frac{3}{2}(\vec{\theta}\cdot\nabla\pi_0)(\vec{B}\cdot\nabla\pi_0)\Biggrb] \,\,,
\end{align}

\ni where we have not, for simplicity, substituted some results obtained in (\ref{dynamics2}) and $\tilde{\cal L}$ is given in Eq. (\ref{10}).  One can see directly that $\pi_0$ is a new field that appears in the Lagrangian and we will discuss it in a moment.

\bigskip


For the second case, the gauge transformations for the Lagrangian in Eq. (\ref{zzZ}) are given by
\baa
\label{zZ2}
\delta_{2nd\:case} A_0 (\vec{x})&=&0  \,\,,\nonumber \\
\delta_{2nd\:case} \vec{A}(\vec{x})&=&-\nabla \epsilon(\vec{x})\,+\,\frac 12 m^2 (\vec{A} \times \vec{\theta}) \epsilon(\vec{x}) \,\,, \nonumber \\
\delta_{2nd\:case} \pi_0 (\vec{x})&=&-\,m^2\,\Big( 1 - \frac 12 \vec{\theta}\cdot \vec{B}\,\Big) \epsilon(\vec{x}) \,\,, \nonumber \\
\delta_{2nd\:case} \vec{\pi}(\vec{x})&=& \frac 12 m^2 \Big[ (\vec{\pi}\times \vec{\theta})\,\epsilon(\vec{x})\,-\,(\vec{\theta}\times \nabla )(A_0 \epsilon(\vec{x}) ) \Big] \,\,,
\eaa

\ni and, as in the previous case, it can be verified after some algebra that the Lagrangian in Eq. (\ref{zzZ}) is invariant under the above transformations.  In this way, we can show that the $\pi_0$ field acts like a Stueckelberg field as it was shown in \cite{Vyt2}.  After rescaling, we can demonstrate that the final Lagrangian is the Stueckelberg one in NC phase space.  In other words, $\pi_0$ helps in order to make the NC Lagrangian invariant.



\section{conclusions}

To construct NC versions of classical systems in theoretical physics is one way to introduce terms or quantities that have a measurement defined at the Planck scale.  It can be considered a semi-classical way since the Planck constant itself was not properly introduced.  However, a NC space-time can also be considered fuzzy and this resulting fuzzy space-time is a path to understand quantum gravity, besides other formulations, of course.

One of the main ingredients of the Standard Model, and since we can consider formulations of noncommutativity as formulations beyond the Standard Model, we can say that to study gauge invariance of NC versions of standard theories is to study gauge invariance beyond the Standard Model.

This is the main target of this work, where we have constructed a gauge invariant version of the NC space-time Proca model.  We believe that it is an interesting result since in its classical (commutative) formulation, the Proca model is not gauge invariant thanks to its mass term.  Since NC formulations do not change the degrees of freedom, to convert this second-class system into a first-class (gauge invariant) one, where the NC parameter terms are present, is a new result in the NC constraint literature.  

With this idea in mind, in this paper we have considered to convert a NC second-class system, the NC Proca model, into a first-class system.  Since we have two second-class constraints, following the GU procedure, we have analyzed two cases where one of the constraints is the chosen first-class one and the second is discarded.  After that we have constructed the first-class Hamiltonian and the respective Lagrangian.  Besides, to obtain the gauge invariance, we have obtained the gauge symmetries.
We have demonstrated precisely that, with the calculated symmetries for both cases, both Hamiltonian and Lagrangian are gauge invariant.  In the second case calculation, a Stueckelberg field was naturally introduced, which helped the construction of the gauge invariant Lagrangian.



\section{Acknowledgments}

\ni The authors thank CNPq (Conselho Nacional de Desenvolvimento Cient\' ifico e Tecnol\'ogico), Brazilian scientific support federal agency, for partial financial support, Grants numbers 302155/2015-5, 302156/2015-1 and 442369/2014-0 and E.M.C.A. thanks the hospitality of Theoretical Physics Department at Federal University of Rio de Janeiro (UFRJ), where part of this work was carried out.  R. L. F. thanks FAPEMIG (Funda\c{c}\~ao de Amparo \'a Pesquisa do Estado de Minas Gerais) for financial support.

\end{document}